\documentclass[final,2p,times,twocolumn]{elsarticle}
\usepackage[utf8]{inputenc}
\usepackage[english]{babel}
\usepackage{lipsum}
\usepackage{amsmath}
\usepackage{amsfonts}
\usepackage{amssymb}
\usepackage{makeidx}
\usepackage{graphicx}
\usepackage[figuresright]{rotating}
\journal{Materials today communication}

\begin{document}

\begin{frontmatter}

\title{A DFT study of mechanical properties of hcp rhenium}
 \author{George S. Manyali } 
\address{$^1$Computational and Theoretical Physics Group, Department of Physical Sciences, Kaimosi Friends University College P.O Box 385-50309, Kaimosi Kenya} 

\fntext[fn1]{gmanyali@kafuco.ac.ke}
\begin{abstract}
In the present paper, hcp Re was investigated in terms of its structural, elastic, mechanical and thermodynamic properties using density-functional theory (DFT). The local density approximation was employed for the exchange-correlation potential together with a spin-orbit coupling. The computed lattice constant was found to be in agreement with the available
experimental and theoretical results. The elastic constants were also calculated and used to determine mechanical properties
like Young's modulus (Y), the shear modulus (G), Poisson's ratio (n) and Vicker's hardness. From thermodynamic investigations, the heat capacity and entropy were also predicted. Although the predicted bulk modulus of Re is comparable to that of diamond, the Vickers hardness was found to be five times less than that of a diamond. Hence, Re is typical solid with high bulk modulus but low Vicker's hardness.

\end{abstract}


\end{frontmatter}
\section{Introduction}
Rhenium (Re) is among very few refractory elements known to exist in a hexagonal-close-packed (hcp) crystal structure contrary to the body-centered-cubic structure shared by Nb, Mo, Ta, and W \cite{de2012first}. While most of refractory elements display the brittle-to-ductile transition, Re does not. Instead, Re features high ductility at low temperature while other elements are known to be brittle at very low temperature, and their toughness become much higher at elevated temperature\cite{carlen1994rhenium,campbell1959availability}. More exciting properties of Re have been uncovered by studying its elasticity.  Generally, mechanical properties describe the response of a material to deformation. Elasticity of a solid can therefore be described in the way in which the atoms in the solid are bonded and how those bonds behave under varying environment\cite{tse2010intrinsic}. Re is known to have  high density of valence electrons\cite{brazhkin2018myths}. Interestingly, diamond has the highest atomic density and the highest density of valence electrons and these parameters are said to be responsible for its highest bulk modulus\cite{brazhkin2018myths}. This implies that Re in principle should have bulk modulus comparable to that of diamond. Is it possible to obtain a material more incompressible than diamond? This is a question of great scientific interest. 

\begin{figure}[h!]
\includegraphics[width=0.4\textwidth]{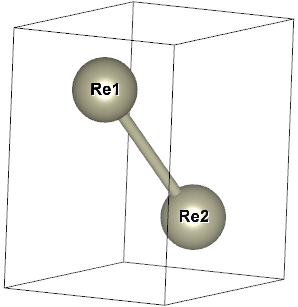}
\caption {(Color on line): The unitcell of hcp Re with two atoms at (0.333333 0.666667 0.250000) and (
0.666667 0.333333 0.750000) positions. }
\label{unit}
\end{figure}

In this article, we report a comprehensive study on mechanical properties of Re and use them as examples to underscore key factors necessary in the search for superhard materials. Re has a simple hexagonal structure (space group P63/mmc) as shown in Figure \ref{unit}.
The elastic tensor is first determined from \textit{ab initio} calculations and derived properties such as bulk modulus, shear modulus, Young’s modulus, Poisson ratio and sound velocity, and Vicker's hardness (Hv) are easily extracted from the elastic tensor. In addition, ductility of Re was predicted using the Pughs ratio.

The paper is organized as follows. In Section 2, the details of first-principles calculations are presented. Section 3 contains the results and their discussion. Finally, in Section 4, we summarize the main conclusions.

\section{Computational details}
The calculations in present work were carried out using Density Functional Theory (DFT) and plane waves basis sets as implemented in the Quantum Espresso package\cite{giannozzi2009quantum,kohn1965self}. The thermo\_pw package \cite{urru2019spin} was used to run several self-consistent and elastic constants calculations. The Local Density Approximation (LDA)\cite{perdew1981self}  exchange-correlation functional were used for all calculations. For spin-orbit coupling, Re.rel-pz-spn-kjpaw\_psl.1.0.0.UPF pseudopotentials was employed while Re.pz-spn-kjpaw\_psl.1.0.0.UPF pseudopotentials was in nonspin orbit coupling calculations. All pseudopotentials were obtained from the pslibrary1.0 maintained by Dal Corso\cite{dal2014pseudopotentials}. The k-point sampling was performed on 16x16x10 grid. The kinetic energy cut-off was set to 60 Ry while the charge density cut-off was set to 400 Ry. The Methfessel-Paxton smearing scheme of width 0.02 Ry was used throughout the calculation.

\section{Results}
\subsection{Structural properties}
The unit cell of Re in hcp structure consists of two atoms. The calculated values for lattice constants are presented in Table \ref{bulkk}. Present data are in good agreement with previous DFT data as well as experiment. The lattice parameter together with isothermal bulk modulus, its deravative and equilibrium total energies were obtain from the Murnaghan equation of state\cite{murnaghan1944compressibility}. The densities of Re calculated with and without SOC are 21.57 and 21.62  g/cm$^3$ respectively. These values are important in calculating the sound wave velocities of Re.
\begin{table}[h!]
\caption{Lattice constant a(a.u), unit cell volume (a.u)$^3$, bulk modulus (K in GPa), derivative of the isothermal bulk modulus (K'), ground-state energy (E0 in Ry), density $\rho$ (in g/cm$^3$) calculated and compared with the experimental and theoretical values of Re.}
\begin{tabular}{lllll}
Property&LDA&LDA+SOC&Others$^a$&Expt$^b$\\\hline
a&5.171   &5.175&5.175 &5.217  \\
c&8.331   &8.337&8.338 &8.425   \\
K& 402.8  &397.5 &372  \\
K'&4.364   &4.410    \\
E0&-1552.9&-1562.9     \\
$\rho$   &21.62&21.57  \\\hline
\end{tabular}\\
$^a$Ref.\cite{urru2019spin}\\
$^b$Ref.\cite{wyckoff1964crystal,kittel1976introduction}\\
\label{bulkk}
\end{table}

\subsection{Elastic properties}
When a crystal is subjected to deformation of a certain magnitude of strain $\epsilon_i$, the restoring forces appear to bring it to the equilibrium configuration. These restoring forces can be described by the stress tensor. In the case of a small strain deforming a material, the stress tensor is proportional to the strain and is given as:
\begin{equation}
\sigma_{i} =\sum_{il} C_{i,l} \epsilon_l 
\end{equation}
where, $C_{i,l}$ is the tensor of elastic constants. The nonzero components of the elastic constant tensor for each crystal point group can be derived using group theory. The form of the elastic constants tensor depends on the Laue class obtained by adding an inversion center to the group operations. In the case of a hexagonal crystal, the form of the elastic constants tensor has five most important independent components. These are: $C_{11}=C_{22}$, $C_{12}$, $C_{13}=C_{23}$, $C_{33}$, $C_{44}=C_{55}$, $C_{66}={1\over 2} (C_{11}-C_{12})$. Hence, the elastic matrix of hexagonal system is given as:
\\
\\
$\begin{pmatrix}
C_{11} &C_{12}& C_{13}&      &     &&\\
       &C_{11}& C_{13}&      &     &&\\
       &      & C_{33}&      &     &&\\
       &      &       &C_{44}&     &  &\\
       &      &       &      &C_{44}&&\\
       &      &       &      &      &C_{66}&\\

\end{pmatrix}
$
\\
\\
The following four necessary and sufficient conditions\cite{bornelastic} for elastic stability in the hexagonal are:
\begin{equation}
C_{11}>|C_{12}|;2C_{11}^2<C_{33}(C_{11}+C_{11});C_{44}>0; 
C_{66}>0
\end{equation}
The elastic constants determined with and without SOC are compared with previous DFT results and experiment as presented in the Table \ref{elastt}. Notice that our calculated values do not vary much from the experimental data. The difference can be accounted for since the DFT data was computed at 0 K. Inclusion of SOC in calculations of elastic constants affects the values with not more than 20~GPa. 
Generally, these single-crystal elastic properties are important input for large scale modeling of other averaged mechanical properties.
\subsection{Bulk modulus and compressibility}
In this work, we related bulk modulus to compressibility. The fundamental definition of isothermal compressibility is the derivative of the equilibrium volume with respect to pressure divided by the volume, calculated at constant temperature as given in Eq.~\ref{kt}:
\begin{equation}
K= - {1 \over V} {d V \over d p}\Bigg|_T.
\label{kt}
\end{equation}
Therefore, isothermal bulk modulus is the inverse of the isothermal compressibility. The calculated values of isothermal bulk modulus and its derivative are presented in Table. \ref{bulkk}. In some cases it might be useful to express the bulk modulus in terms of the elastic constants using the Voigt and Reuss averages of the bulk modulus. The Voigt average provides an upper bound on the elastic moduli of an untextured polycrystalline material whereas the Reuss average provides a lower bound\cite{hill1952elastic}. The averaged values of bulk moduli obtained with and without SOC are presented in Table \ref{elastt}. The K value presented in Table \ref{bulkk} and B value determined from elastic constant differ with less than 3GPa. This confirms that both approximation methods produces similar results. The maximum value of B recorded in this work was about 405 GPa. The value rivals the bulk modulus of cubic boron nitride as well,implying that rhenium can be considered to an ultrahard material due to its high resistance to volume change. 

\begin{table}[h!]
\caption{Elastic constants C$_{ij}$ (in GPa), Bulk modulus B (in GPa), derivative of bulk modulus (B') Shear modulus G (in GPa), Young's modulus Y  (in GPa),  poisson's Ratio $\nu$, sound velocities V\_p and V\_s (in m/s), Debye temperature $\theta_D$ (in K), ( B/G ratio, and Vicker's Hardness (in GPa).}
\begin{tabular}{llllll}\hline
       &LDA&LDA&Others$^a$&Expt$^b$.\\
        &    &+SOC&         &       \\\hline
C$_{11}$&694.1   &692.8&605,672    &616,634  \\
C$_{33}$&733.5   &739.1&650,740    &683,701   \\
C$_{12}$&270.5   &272.3&235,309    &273,266   \\
C$_{13}$&247.1   &233.2&195,252    &206,202   \\
C$_{44}$&176.5   &179.1&175,176    &161,169    \\
B       &405.6   &400.2&390 &  \\
Y       &519.8   &524.0&456  & \\
G       &202.2   &204.4&175   &\\
$\nu$   &0.286   &0.281&   \\
G/B     &0.498   &0.510&    \\
Hv      & 16.7   &17.4 & &\\
V$_P$& 5302 &5204 &  \\
V$_B$& 4331 &4307  &\\
V$_G$& 3058 &3078 &\\
$\theta_D$&417&419 & \\\hline
\end{tabular}\\
$^a$Ref.\cite{steinle1999first,de2012first,lv2012elastic}\\
$^b$Ref.\cite{katahara1979pressure,simmons1965single}\\
\label{elastt}
\end{table}

\subsection{Vicker's hardness, Shear and Young modulus}
The shear modulus (G), describes the calculated plastic twist of a material, while Young's modulus (Y), presents the strength of the material. Both values of the shear and Young's modulus have been obtained from the Voigt–Reuss–Hill and recorded in Table \ref{elastt}. Previous studies have demonstrated that there is intrinsic correlation between hardness and shear modulus. Hence, shear modulus does provide a correct assesment of hardness materials than the bulk modulus\cite{liu2017elastic,manyali2013ab}. 

Hardness is the ability of a material to resist elastic deformation, plastic deformation, or failure under external force. 
This concept was well illustrated in 1954 when Pugh proposed a relation between the elastic and plastic properties of pure polycrystalline metals\cite{pugh1954xcii}. The Pugh’s modulus ratio  G/B is the basis of the so called Chen model \cite{chen2011modeling} that has been widely used to predict  Vicker's hardness of a variety of crystalline  metals, insulators and semiconductors. The Chen model is given as:
\begin{equation}
Hv =2(\frac{G^3}{B^2})^{0.585}-3.
\label{hvv}
\end{equation}
Where G and B are shear and bulk modulus respectively. The Vickers hardness of rhenium element is presented in Table \ref{elastt}. The low value of Hv is an indicator of weak covalent bonds which is presented by a large value of Poisons ratio ($\nu$). Compared to well known hard materials such as diamond and cubic boron nitride, rhenium element does not fall in the same category of ultrahard materials (Hv $>$ 80~GPa). 
Furthermore, the Pugh’s modulus ratio represents a good criterion to identify the brittleness and ductility of materials. The higher the value of G/B is, the more brittle the materials would be. Diamond a hardest known brittle material has G/B ratio of about 1.2, yet rhenium has a G/B ratio of about 0.51. This is a clear indication that rhenium is more ductile than diamond.

 \subsection{Thermal properties}
The Debye temperature ($\theta_D$),describe the thermal characteristics of the rhenium element. We report a value of $\theta_D$ as 419~K and 417~K calculated with and without SOC respectively. The eq. \ref{deby} show how $\theta_D$ was determined:
\begin{equation}
\theta_D=\frac{h}{k} \left[\frac{3n}{4\pi}\left(\frac{N_A\rho}{M}\right)    \right]^{-1/3}V_m
\label{deby}
\end{equation}
where $h$ is Planck’s constants, k is Boltzmann’s constant, N$_A$ is Avogadro’s number, n is the number of atoms in the molecule, M is the molecular weight, and $\rho$ is the density. The average wave velocity V$_m$ is approximately calculated from 
\begin{equation}
V_m=\left[\frac{1}{3} \left(\frac{2}{V_s^3}+\frac{1}{V_p^3}  \right)  \right]^{-1/3}
\label{vmm}
\end{equation}

where V$_P$ and V$_G$ are the compressional and shear wave velocities, respectively, which can be obtained from Navier’s equation \cite{anderson1963simplified}.  

\begin{equation}
V_P=\sqrt{\left( B + \frac{4}{3}G \right)\frac{1}{\rho}}
\label{vppp}
\end{equation}

\begin{equation}
V_G=\sqrt{ \frac{G}{\rho}}
\label{vSS}
\end{equation}

 The bulk sound velocities are given as;
\begin{equation}
V_B=\sqrt{ \frac{B}{\rho}} .
\label{vbS}
\end{equation} 
Sound velocities for rhenium element are presented in Table \ref{elastt}. The compressional wave velocities are above 5200 m/s while the shear wave velocities are about 3000m/s. The bulk wave velocities behaves like an average of both compressional and shear wave velocities.

 \subsection{Thermodynamic properties}

\texttt{\begin{figure}[h!]
  (a)\includegraphics[width=0.34\textwidth]{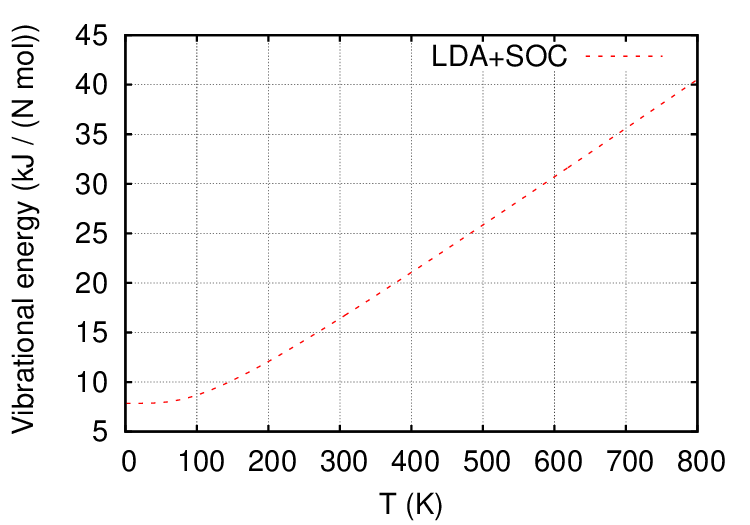}
  (b)\includegraphics[width=0.34\textwidth]{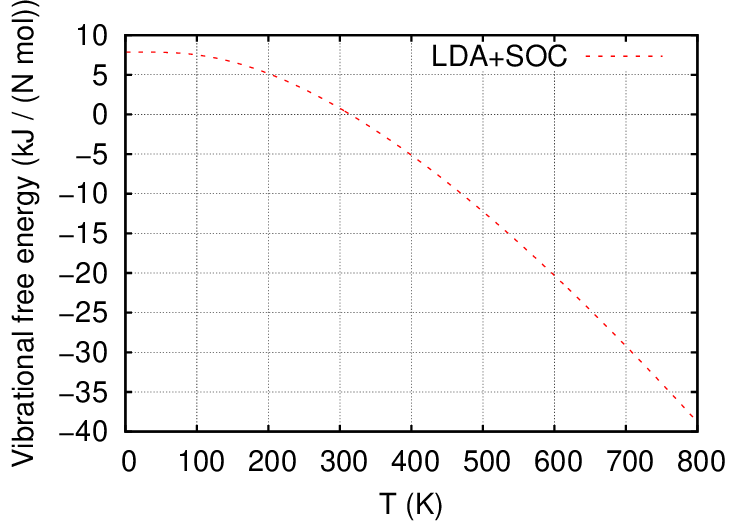}
  (c)\includegraphics[width=0.34\textwidth]{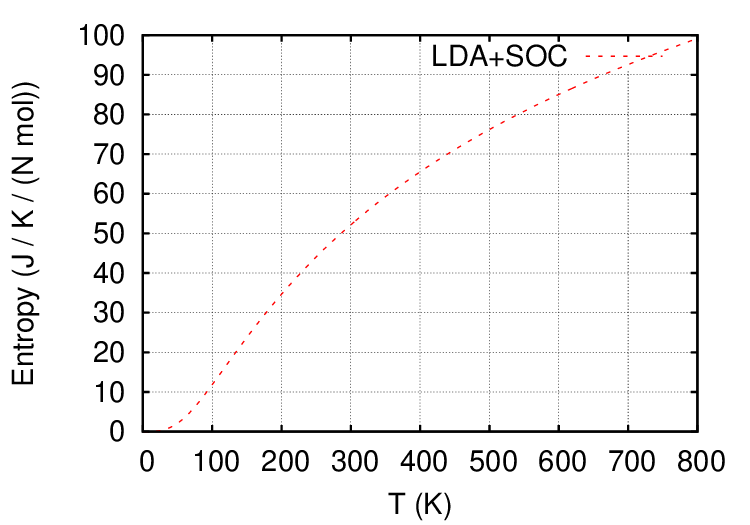}
  (d)\includegraphics[width=0.34\textwidth]{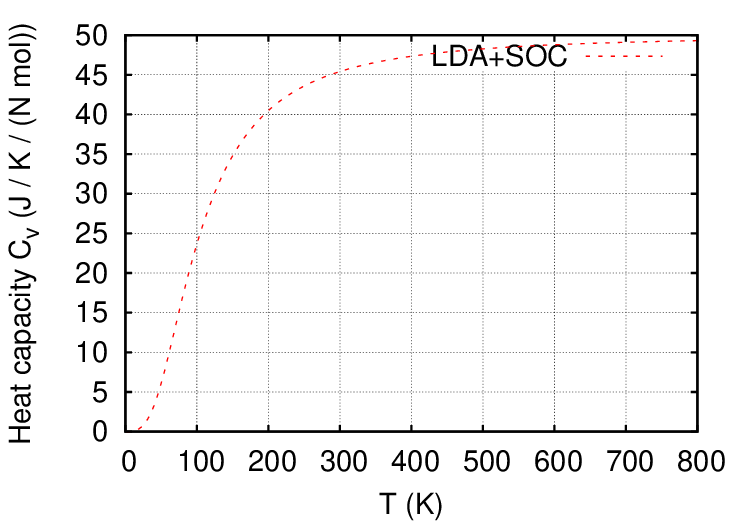}
  \caption {(Color on line): (a) Debye vibration energy, (b) Debye vibrational free energy, (c) Debye entropy (d)Debye heat capacity.}
 \label{figure:THERMO}%
 \end{figure}}
In Figure~\ref{figure:THERMO}, we demonstrate how different properties of materials depend on temperature. The vibrational internal energy, vibrational free energy, entropy and heat capacity for hcp Re, show different behaviour at low and high temperatures. The vibrational internal energies increases with increase in temperature as depicted in Figure~\ref{figure:THERMO}a. Furthermore, the vibrational free energy which is often  associated with the Helmholtz free energy is shown in Figure~\ref{figure:THERMO}b. As temperature increases, the free energy decrease. Figure~\ref{figure:THERMO}c, show entropy dependence on temperature. Entropy is a thermodynamic property that is very sensitive to temperature change. It measures the state of disorder of a molecular system. Figure~\ref{figure:THERMO}d, show the temperature dependence of heat capacity at constant volume. Above 400K, Re obeys the Dulong and Petit law.

 \section{Conclusion}

In closing, we have carried out a comprehensive study of mechanical properties of hcp Re. We begun by calculating the elastic tensor from which we obtained the five independent elastic constants. The averaged bulk modulus, shear modulus, Young's modulus, poison ratio, debye temperature and sound wave velocities were easily obtained from the elastic constants. Other thermal properties i.e heat capacity, entropy and vibrational internal energies were also obtained. Overall, we found good agreement between our calculated values and experimental values. The question on possibility of having materials that rival the bulk modulus of diamond was answered by showing, Re has a bulk modulus of about 400~GPa which is close to that of diamond. But as to whether there is a linear correlation between hardness and bulk modulus, we found that for the case of Re, the high bulk modulus does not translate to high Vicker's hardness. Re was predicted to have  Vicker's hardnes of about 17~GPa which is almost 5 times less than that of diamond. Hence, Re is a typical material with high bulk modulus but low Vicker's hardnes. We also demonstrated through Pugh's modulus ratio that Re is a ductile material.

 \section*{Acknowledgment}
Work was supported by African Laser Centre Research Grant No. LHIN500 Task ALC-R005. Computational resources were provided by CHPC, South Africa.

\bibliographystyle{elsarticle-num} 
\bibliography{ref}
\end{document}